\documentclass[shortnote]{jpsj3}
\usepackage{txfonts}

\title{Synthesis of New Layered Oxypnictides Sr$_2$CrO$_2$(FeAs)$_2$}

\author{
\name{Naoya \surname{Eguchi}}$^1$\thanks{E-mail: naoya{\_}eguchi@phys.sc.niigata-u.ac.jp}, 
\name{Fumihiro \surname{Ishikawa}}$^2$, 
\name{Michihiro \surname{Kodama}}$^1$, 
\name{Takeshi \surname{Wakabayashi}}$^1$, 
\name{Atsuko \surname{Nakayama}}$^3$, 
\name{Ayako \surname{Ohmura}}$^3$, 
and \name{Yuh \surname{Yamada}}$^2$}
\inst{
$^1$Graduate School of Science and Technology, Niigata University, Niigata 950-2181, Japan\\
$^2$Department of Physics, Niigata University, Niigata 950-2181, Japan\\
$^3$Center for Transdisciplinary Research, Niigata University, Niigata 950-2181, Japan} 

\kword{Fe-Based superconductor, layered oxypnictides}
\begin{document}
\maketitle

Since the discovery of LaFeAs(O, F) superconductor \cite{LaFeAsO}, 
a large number of iron-oxypnictides with superconductivity have been investigated.
In the oxypnictide superconductors,
blocking layers, which exist between FeAs layers, characterize series of compounds;
LaO block in LaFeAs(O, F), for example.
A series of oxypnictides having perovskite-type blocking layers is attractive because of their highly two dimensional crystal structure.
In fact, Sr$_4$V$_2$O$_6$(FeAs)$_2$ shows superconducting transition at 37 K \cite{Sr42622} under ambient pressure without doping,
while most of FeAs-based superconductors needs doping or applying pressure.

As reviewed by Ozawa and Kauzlarich\cite{review}, several layered oxypnictides structures are considerable candidates as a new superconductor.
Sr$_2$Mn$_3$As$_2$O$_2$-type oxypnictide is one of the layered oxypnictides introduced in the review. 
Nath \textit{et al.} reported their attempt to synthesize Sr$_2$Mn$_3$As$_2$O$_2$-type oxypnictides and succeed to prepare several kinds of polycrystalline samples
with MnAs layers\cite{Sr2322}.
They also reported that they could not obtain the single phase samples with Fe-substituted compositions, for example, Sr$_2$Fe$_3$As$_2$O$_2$.

\begin{figure}[htbp]
\begin{center}
	\includegraphics[width=8.2cm]{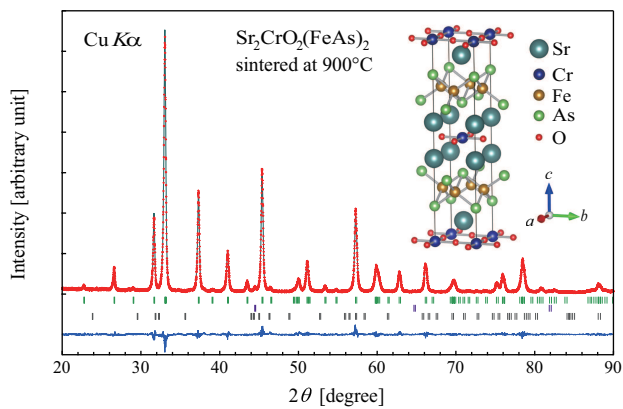}
	\caption{(Color online) X-ray powder diffraction pattern and Rietveld fit for Sr$_2$CrO$_2$(FeAs)$_2$ sintered at 900$^\circ$C. 
	Small amount of Fe-Cr alloy ($<$1\%) 
	and SrFe$_2$As$_2$ ($<$1\%) were included as minor impurity phases.
	Dots and 
	solid line represent the observed and calculated intensities, respectively. 
	Upper ticks below the profile mark the positions of the reflections from the main phase.
	Middle and lower ticks indicate two impurity phases, Fe-Cr alloy and SrFe$_2$As$_2$, respectively. 
	Solid line at the bottom shows the residual error. Inset shows the crystal structure of  Sr$_2$CrO$_2$(FeAs)$_2$ drawn by VESTA program\cite{VESTA}. }
	\label{x-ray}
\end{center}
\end{figure}

Recently, we successfully synthesized Fe-substituted Sr$_2$CrO$_2$(FeAs)$_2$, which is isostructural with Sr$_2$Mn$_3$As$_2$O$_2$
except for Cr-substitution of Mn-site in the blocking layer.
The crystal structure of  Sr$_2$CrO$_2$(FeAs)$_2$ is shown in the inset of Fig. \ref{x-ray}.
The FeAs layers are separated by two Sr-layer and CrO$_2$-layer according to the stacking order of (Sr)(CrO$_2$)(Sr).
In this paper, we represent our results of the crystal structure determined by x-ray powder diffraction and the physical properties
of Sr$_2$CrO$_2$(FeAs)$_2$.
Briefly, Sr$_2$CrO$_2$(FeAs)$_2$ does not show superconductivity in analogy with Sr$_4$Cr$_2$O$_6$(FeAs)$_2$\cite{Sr42622Cr,Sr42622Cr_Ogino}.

Polycrystalline	sample of  Sr$_2$CrO$_2$(FeAs)$_2$ is
prepared by solid-state-reaction techniques using SrO (98\% pure),
FeAs (99.5\% pure), 	
Cr (99.9\% pure), 
as starting materials.
 Composition of starting mixture is nominal.
A pellet of the ground mixtures of the starting materials wrapped with tantalum foil was sealed inside double quartz tubes with argon gas.
All the sample handling was carried out inside an Ar-filled glove box.
The samples were heated to 800$^\circ$C, 900$^\circ$C and 1000$^\circ$C at a rate of 50$^\circ$C/h, held there for 96 h, respectively. 
Each sample mainly consists of Sr$_2$CrO$_2$(FeAs)$_2$ phase, which was confirmed by powder x-ray diffraction.
However, the samples sintered at 800$^\circ$C and 1000$^\circ$C also contain FeAs phases, clearly.
In addition, the sample sintered at 800$^\circ$C has a large amount of other impurity phases such as SrFe$_2$As$_2$ and Sr$_4$Cr$_2$O$_6$(FeAs)$_2$.
Thus, below in this paper, we only describe the sample of Sr$_2$CrO$_2$(FeAs)$_2$ sintered at 900$^\circ$C.

Powder x-ray diffraction patterns were measured
using Cu$K\alpha$ radiation in the 2$\theta$ range of 20-90$^\circ$ (RINT-2000, Rigaku, Japan) .
Rietveld refinements were made by the analysis program RIETAN-FP\cite{rietan}.
In Fig. \ref{x-ray}, the x-ray diffraction pattern for the sample sintered at 900$^\circ$C is shown with the results of Rietveld refinement profile.
 Sr$_2$CrO$_2$(FeAs)$_2$ is obtained as the main phase with a small amount of impurities,
 such as Fe$_{1-\textit{x}}$Cr$_{\textit{x}}$ and SrFe$_2$As$_2$.
According to the results of multi-phases peak-fitting with RIETAN-FP, total amount of impurity phases are less than 2\%.
Refined results by Rietveld analysis are summarized in Table \ref{results} .
The space group of the compound is \textit{I}4/\textit{mmm} and the lattice constants are 
$a$ = 3.9948(1) $\AA$ and 
$c$ = 18.447(1) $\AA$. 
The errors in the lattice parameters are the standard errors evaluated by Rietveld analysis.

\begin{table}[htbp]
	\begin{center}
	\caption{Structure parameters for Sr$_2$CrO$_2$(FeAs)$_2$
	refined from powder x-ray diffraction data with the space group of $I$4/${mmm}$;
	 $R_{wp}$ = 8.65\%, $R_{p}$ = 6.33\%, $S$ = 1.23.}\label{results}
	\tabcolsep=6mm
		\begin{tabular}{ccccc}
		Atom		& 	Site	&	 x	&	 y 	&	z	\\ \hline \hline
		Sr		&	$4e$	&	0	&	0	&	0.4098(1)	\\
		Cr		&	$2a$	&	0	&	0	&	0	\\
		O		&	$4c$	&	0	&	0.5	&	0	\\
		Fe		&	$4d$	&	0	&	0.5	&	0.25	\\
		As		&	$4e$	&	0	&	0	&	0.1753(1)	\\ \hline
		\end{tabular}
	\end{center}
\end{table}

\begin{figure}[htbp]
\begin{center}
	\includegraphics[width=7.2cm]{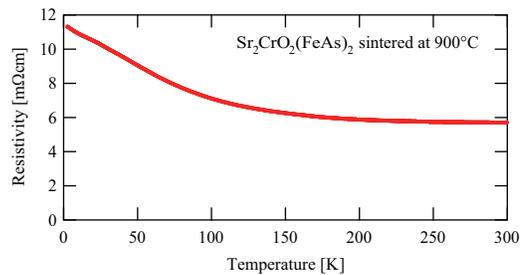}
	\caption{(Color online) Temperature dependence of the resistivity for Sr$_2$CrO$_2$(FeAs)$_2$. }
	\label{resistivity}
\end{center}
\end{figure}

The electrical resistivity was measured by a conventional four-probe method with a Gifford-McMahon cryocooler between 3 and 300 K.
Temperature dependence of the resistivity for Sr$_2$CrO$_2$(FeAs)$_2$ is shown in Fig. \ref{resistivity}.
Over whole measured temperature range, Sr$_2$CrO$_2$(FeAs)$_2$ shows semiconductor-like behavior.
Additionally, the value of the resistivity is smaller than that of Sr$_4$Cr$_2$O$_6$(FeAs)$_2$, which also shows nonmetallic behavior with the resistivity of about 60 m$\Omega$cm at room temperature\cite{Sr42622Cr}.
While many kinds of  iron-oxypnictide superconductor and related compounds show the anomaly due to structural and magnetic transition,
Sr$_2$CrO$_2$(FeAs)$_2$ did not show any trace of such transition in the resistivity variation with temperature.

\begin{figure}[htbp]
\begin{center}
	\includegraphics[width=7.2cm]{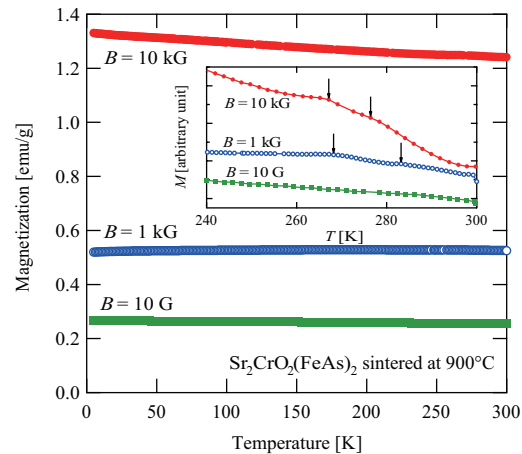}
	\caption{(Color online) Temperature dependence of the magnetization for Sr$_2$CrO$_2$(FeAs)$_2$ at several magnetic fields.
	 Inset shows enlarged view near room temperature.
	Note that each curve is offset and plotted in arbitrary unit in the inset.
	Arrows indicate kinks in the curves.}
	\label{MT}
\end{center}
\end{figure}

The magnetization up to 10 kG was measured using with a SQUID magnetometer (MPMS of Quantum Design Co.). 
Temperature dependence of the magnetization for Sr$_2$CrO$_2$(FeAs)$_2$ at several magnetic fields is shown in Fig. \ref{MT}.
As shown in this figure, 
variation in magnetization is very small and shows no Curie-Weiss-like behavior.
This is in contrast to the previously reported results for Sr$_4$Cr$_2$O$_6$(FeAs)$_2$ \cite{Sr42622Cr,Sr42622Cr_Ogino}.
Tegel \textit{et al.} reported antiferromagnetic order of Cr$^{3+}$ moment \cite{Sr42622Cr}.
The inset in Fig. \ref{MT} shows the magnified view of curves between 240 K and 300 K with vertical offset.
In the magnetic field of 1  and 10 kG, small kinks appeared in the curves; the origin of these kinks is not yet clear.
Additionally, temperature dependence at low temperature at 10 G does not show any trace of the Meissner effect 
even under high pressure up to 1 GPa (not shown here).

\begin{figure}[htbp]
\begin{center}
	\includegraphics[width=7.2cm]{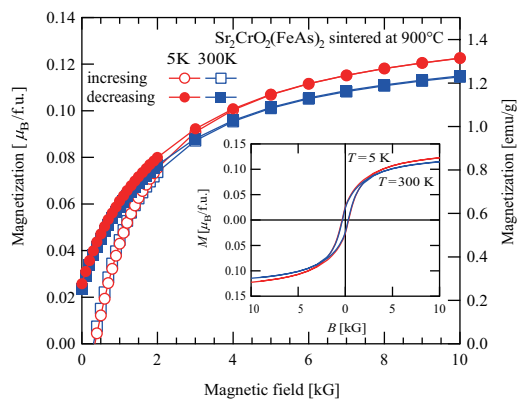}
	\caption{(Color online) Magnetization curves for Sr$_2$CrO$_2$(FeAs)$_2$ at 5 K and 300 K. Inset shows the hysteresis loops at 5 K and 300 K.}
	\label{MH}
\end{center}
\end{figure}

As shown in the magnetization curves in Fig. \ref{MH},
the present sample shows apparent ferromagnetic behavior even at room temperature.
With increasing temperature, magnetization slightly decreases, but coercive force is  almost same between 5 K and 300 K.
These results suggest that the present sample contains some ferromagnetic phases having higher Curie temperature than room temperature.
This ferromagnetic behavior is probably due to small amount of the Fe-Cr alloy which exists as an impurity phase.
We may estimate that the amount of the Fe-Cr alloy phase is about 1\% using our magnetization value and the reported  Fe-Cr alloy magnetization.
For example, bulk magnetization value of Fe$_{65}$Cr$_{35}$ was experimentally determined to be about 138 emu/g \cite{FeCr}.
This estimated content of Fe-Cr alloy phase from magnetization data is consistent with the Rietveld refinement result as mentioned above.

In conclusion,
we successfully prepared new layered iron oxypnictide Sr$_2$CrO$_2$(FeAs)$_2$, which has no superconducting transition down to 3 K.
This is the first report on the Fe-substituted Sr$_2$Mn$_3$As$_2$O$_2$-type oxypnictides to our knowledge.
However, our present sample contains the strong ferromagnetic Fe-Cr alloy as impurity.
Purification is required to investigate intrinsic magnetic properties for Sr$_2$CrO$_2$(FeAs)$_2$.
Moreover, we consider this compound can form a new family of the oxypnictide superconductor.
Doping to the compound or substitution of Cr by other element probably lead to the superconducting phase.
At this time, applying pressure to this compound does not cause the superconducting phase.
Further study of this series is being undertaken.

\begin{acknowledgment}

This research was supported by 
JSPS KAKENHI Grant Number 23684024.
\end{acknowledgment}

\end{document}